# Statistical modelling under differential privacy constraints: A case study in fine-scale geographical analysis with Australian Bureau of Statistics TableBuilder data

Ewan Cameron[a,b,c*]

[a] *Geospatial Health and Development Team, Telethon Kids Institute, Perth, Australia*

[b] *School of Public Health, Curtin University, Perth, Australia*

[c] *Stan Perron Foundation Fellow*

[*] Corresponding author (ewan.cameron@telethonkids.org.au)

# Statistical modelling under differential privacy constraints: A case study in fine-scale geographical analysis with Australian Bureau of Statistics TableBuilder data

Guided by the principles of differential privacy protection the Australian Bureau of Statistics modifies the data summaries from the Australian Census provided through TableBuilder to researchers at approved institutions. This modification algorithm includes the injection of a small degree of artificial noise to every non-zero cell count followed by the suppression of very small cell counts to zero. Researchers working with small area TableBuilder outputs with a high suppression fraction have proposed various algorithmic solutions to reconciling these with less suppressed outputs from larger enclosing areas. Here we propose that a Bayesian, likelihood-based statistical approach in which the perturbation algorithm itself is explicitly represented is well suited to analyses with such randomly perturbed data. Using both real (TableBuilder) and mock datasets representing dwelling classifications in the Perth Greater Capital City Area we demonstrate the feasibility and utility of multi-scale Bayesian reconstruction of modified cell counts in a spatial setting.

**Introduction**

Data summaries from the Australian Census are made available by the Australia Bureau of Statistics (ABS) to researchers at approved institutions through the TableBuilder platform[1]. Users with permission to run queries at the TableBuilder Pro tier can request counts of dwellings, households or persons stratified in up to three 'dimensions' (rows, columns and wafers) along selected Census attributes, such as place of usual residence, age, sex, household size, or household income. Spatial aggregation scales may be chosen from nested Australian Statistical Geography Standard units[2], including Mesh Blocks (containing around 30 to 60 dwellings in most instances), SA1s (around 200 to 800 people) and SA2s (3,000 to 25,000 people). Example applications of TableBuilder outputs for research in spatial epidemiology and geography abound in the literature, such as the identification of gay and lesbian population clusters ('gayborhoods') by postcode (Callander, Mooney-Somers et al. 2020), the estimation of associations between socio-economic index and migrant population densities at 'suburb' (SA2) level (Colic-Peisker and Peisker 2022), and the construction of virtual populations at SA1 level for a mechanistic model of influenza (Zachreson, Fair et al. 2018).

---

[1] https://www.abs.gov.au/statistics/microdata-tablebuilder/tablebuilder

To preserve the confidentially of individual records in the Australian Census, the ABS have developed a methodology for *perturbing* the true counts by Census attribute, such that those returned by TableBuilder are in fact noisy, censored versions of the originals (Chipperfield, Gow et al. 2016). Although the early development of this algorithm was not shaped explicitly by differential privacy theory, it has subsequently been analysed through this lens and found to be broadly consistent with its principles and aims (Rinott, O'Keefe et al. 2018, Bailie and Chien 2019). Namely, to effectively share information about groups of individuals without exposing the individual data to more than a tiny risk of probabilistic de-identification. As Rinott, O'Keefe et al. (2018) note: "*It is a property of differential privacy that the confidentiality protection guarantee does not rely on hiding the parameters of the perturbation. … [hence] knowledge of the mechanism allows the user to take the perturbation distribution into account in their analysis for data independent algorithms*". A number of publications from the ABS give insight into the nature of the perturbation methodology applied within TableBuilder , although in fact none of these documents provide sufficient information from which to perfectly reconstruct the noise generation process (Chipperfield, Gow et al. 2016, Giessing 2016).

---

[2] https://www.abs.gov.au/statistics/statistical-geography/australian-statistical-geography-standard-asgs

Two recent studies have presented algorithmic methodologies to facilitate geospatial analyses of sparse Tablebuilder outputs at fine spatial scales.  In *Australian Geographer*, Kok, Tuson et al. (2022) consider the problem of mapping incidence rates of high risk foot hospitalisations amongst the Indigenous population of Perth.  They demonstrate that the suppression of population counts for Indigenous residents filtered by age group at the SA1 scale creates two problems for epidemiological analyses of this nature: (1) some SA1 areas with attributed hospital address records have zero population output from TableBuilder (yielding an improper zero denominator for the crude incidence rate), and (2) the aggregations of SA1 TableBuilder population totals within their enclosing SA2 boundaries are in many cases markedly below the equivalent TableBuilder outputs returned on querying the SA2 populations totals directly. With these issues in mind the authors propose a 'map overlay technique' in which many separate incidence 'hotspot' maps are created for different aggregations of the SA1 units, from which a fine-scale product is subsequently synthesised using AZTool.

In *Nature Scientific Data*, Fair, Zachreson et al. (2019) consider the problem of reconstructing the commuter networks of Greater Sydney using TableBuilder outputs of counts over place of residence and place of work (POW) destination zone (DZN) pairs. The authors demonstrate the systematic impacts of low count suppression on the network structures built using the outputs available at various spatial scales (both for origins and destinations).  They further propose an algorithm for simulating a reconstructed network of SA1 to POW DZN connections that matches approximately the SA2 to POW DZN totals.  Here random out-edges from each SA1 are proposed from a normalised distribution of edge weights given the residential population of that SA1. Candidate in-edges to POW DZN are then proposed considering SA2 to POW DZN pair totals, SA2 to POW SA2 pair totals and the associated POW DZN totals (independent of SA2 origin) from TableBuilder. It is noted that no new SA2 to POW SA2 connections will be added by this method.

Common to both the algorithms mentioned above is that neither explicitly models the perturbation process and neither employs a likelihood-based statistical formulation of the problem. While both methods involve a degree of randomisation within their algorithms, neither returns a solution with associated confidence or credible intervals on the reconstruction.

As a technical aside before proceeding it is also worth noting that in both studies there is emphasis placed on the previously available 2011 Census TableBuilder outputs as a ground-truth for that year. Prior to the 2016 Census the perturbation algorithm used in ABS TableBuilder included an additional step (after noise injection and suppression) in which the returned counts from finer scale units were rebalanced for consistency with the reported totals in their enclosing coarser scale units. While the mechanism for this adjustment is (to our knowledge) not explicitly reported, the description provided on the ABS website[3] gives no reason to suppose that this step returned counts to suppressed cells. A reasonable assumption might be that the process applied was akin to the simple 'raking' procedure used by the Institute for Health Metrics Evaluation to ensure a one-death-one-cause balance across competing disease models (Foreman, Lozano et al. 2012). For this reason we do not consider the original 2011 TableBuilder outputs to be a ground

---

[3] https://www.abs.gov.au/ausstats/abs@.nsf/Lookup/2916.0main+features252016

truth for the purpose of the present analysis.

In this manuscript we demonstrate a likelihood-based, Bayesian approach to the joint estimation of counts across multiple spatial units from TableBuilder outputs. We propose a generative approximation to the unknown (or 'incompletely known') perturbation algorithm deployed in TableBuilder and describe a spatial prior-structure and random-walk MCMC sampling scheme for posterior simulation under this likelihood function. The nature of the model-based reconstructions and performance of the posterior credible intervals are examined using both real (TableBuilder) and mock datasets. Both the well-specified case (in which the mock data are generated under the same perturbation model assumed by the likelihood) and a misspecified case (in which an alternative noise model is used to generate the mock data) are considered. Posterior predictive checks are demonstrated as a means of model scrutiny in this setting and various avenues for further research are outlined.

**Methods**

***Perturbation Model***

High level details of the perturbation procedure applied to ABS TableBuilder outputs available on the ABS website are as follows:

- "As perturbation is applied *independent* of the size of a count, any individual count or total in Census data products will be no more than a very small number away from the unperturbed value."[4]

---

[4] https://www.abs.gov.au/ausstats/abs@.nsf/Latestproducts/2916.0Main%20Features302016?opendocument&tabname=Summary&prodno=2916.0&issue=2016&num=&view=

- "Perturbation includes the *suppression* of small counts so individual information cannot be determined. This is why you'll never see counts of 1 or 2 in Census output."[5]
- "Perturbation is applied across all *non-zero* cells in a table, including the totals cells."[6]

Hence, we propose a minimal perturbation model, $P(c_{\text{output}}|c_{\text{true}})$, for the output cell count, $c_{\text{output}}$, as:

$$c_{\text{output}} | \{c_{\text{true}} > 0\} = I(c_{\text{intermediate}} > 2) \times c_{\text{intermediate}}$$

with $c_{\text{intermediate}} \sim \text{ZeroTruncatedDiscreteNormal}(c_{\text{true}}, \sigma_{\text{err}}^2)$ and

$$c_{\text{output}} | \{c_{\text{true}} = 0\} = 0$$

where $\sigma_{\text{err}}$ represents the (unknown) standard deviation of the perturbation process for non-zero cells, and $I(c_{\text{intermediate}} > 2)$ represents the suppression step.

Published descriptions of the perturbation modelling techniques developed by the ABS, however, indicate that the actual perturbation algorithm is much more complicated than our minimal version. Crucially, the error perturbations are not drawn independently for every cell in every possible query, but are pre-compiled from a pseudo-random number table generated against the individual micro-census records of persons, families and households (Marley and Leaver 2011, Thompson, Broadfoot et al. 2013). The scale of the applied perturbations may also be varied on a per-query or per-record basis according to the sensitivity of information at-risk of disclosure (Marley and Leaver 2011). Interestingly, there are also a number of additional complications to the perturbation algorithm, such as the use of asymmetrical random noise models, that have been flagged as 'under investigation' on the ABS website[7].

---

[5] Ibid.

[6] https://www.abs.gov.au/statistics/microdata-tablebuilder/tablebuilder/confidentiality-and-relative-standard-error

The framework under which inference with the minimal perturbation model proposed here will thus be considered is that of Bayesian modelling under misspecification (Gelman and Shalizi 2013). To this end we must first make clear the purpose of our modelling exercise, which we will attempt to carry through to the design of our model and our investigations of its performance. Three principal objectives that reflect the context given in the Introduction above are now proposed:

1. the modelling should return estimated count maps at each areal scale that are mutually consistent, meaning that the counts imputed in finer scale areal units sum to the imputed totals of their coarser enclosing areal units;

2. the modelling should return probabilistic statements regarding the confidence attached to the total counts in each areal unit and the probability that a zero cell is structural (i.e., a true zero, rather than a small non-zero count randomly suppressed to zero); and

3. the modelling should permit the introduction of ancillary data and/or statistical constraints to ensure that the estimated count maps acknowledge contextual factors relevant to the scientific analysis for which they are being reconstructed.

These objectives speak directly to the concerns that have been raised by researchers working with TableBuilder outputs following the removal of the post-perturbation additivity adjustments from 2016 onwards. Namely, that the suppression of low cell counts leads to an apparent under-counting when comparing between spatial aggregations at different areal scales and that in some instances zero counts are returned

---

[7] https://www.abs.gov.au/statistics/research/methodological-news-jun-2022

where ancillary information suggests otherwise (Fair, Zachreson et al. 2019, Kok, Tuson et al. 2022).

An alternative perturbation model that achieves a similar effect to the above model in aggregate, but which operates at the individual record level is as follows. For each record, $r = 1, \ldots, N_{\text{ind}}$, in the dataset, a 'sensitivity ranking', $s_r$, is drawn from the (continuous) uniform distribution, $s_r \sim \text{Uniform}(0,1)$, along with a sensitivity-scaled error term, $e_r$, from the Normal distribution, $e_r \sim \text{Normal}(0, (2s_r)\sigma_{\text{err}}^2/\sqrt{5})$. For every cell, in any query, the true count is modified to $c_{\text{intermediate}}$ by adding the error terms of the top 5 individuals in that cell by sensitivity ranking (or all individuals if there are fewer than 5), i.e., $c_{\text{intermediate}} = c_{\text{true}} + \sum_{r \in \text{rank}(s_r)[1:5]} e_r$. Rounding to the nearest integer, followed by suppression of all values of 2 and below then produces the final $c_{\text{output}}$. The significance of the $\sqrt{5}$ term above is that for cells containing 5 or more records the expected standard deviation of the summed error terms matches the proposed error scale, $\sigma_{\text{err}}$.

Under this perturbation model, which we will refer to henceforth as 'the individual record perturbation model', there will be a correlation in the outputs of small cells with commonality amongst their sets of most sensitive members. The sampling distribution for this model is intractable for likelihood-based analyses of aggregate counts. For both models we will suppose here a fixed value of $\sigma_{\text{err}} = 2$. The reason for this choice is that if one compares the output row totals against row sums for combinations of TableBuilder categories and areal units in which all entries are far from zero (e.g. counts by sex at SA2 level), the empirical standard deviation is always close to $2 \times \sqrt{M}$ where $M$ is the number of columns including the row total itself, suggesting a typical error scale for large population cells of around 2.

*Bayesian Hierarchical Model*

We suppose here that the input dataset for modelling consists of TableBuilder outputs of counts, $c_{\text{output}}^{i,j,k}$, for a categorical variable with $i = 1, \ldots, M-1$ classes plus the row totals at $i = M$ for each of $j = 1, \ldots, N_k$ areal units at $k = 1, \ldots, A$ nested levels. For ease of explanation, we will make the latter concrete with $k = 1$ being Mesh Blocks (smallest), $k = 2$ being SA1 areas and $k = 3$ being SA2 areas (largest). The real TableBuilder dataset we use in our subsequent experiments is restricted to the Perth Greater Capital City region (Census 2021) for which $N_1 = 27{,}112$, $N_2 = 4{,}822$ and $N_3 = 185$. Moreover, we examine counts across the four classes (i.e., $M = 5$) of 1-digit level Family Household Composition (HCFMD): "Multiple Family Household", "One Family Household", "Other Household" and "Not Applicable".

Conditioning our minimal perturbation model on a set of latent (i.e., unknown) 'true' counts (the target of our inference), $\{c_{\text{true}}^{i,j,k}\}$, creates a likelihood function, $L(\{c_{\text{true}}^{i,j,k}\}) = \prod_{i,j,k} P(c_{\text{output}}^{i,j,k} | c_{\text{true}}^{i,j,k})$. An improper prior for $\{c_{\text{true}}^{i,j,k}\}$ in the spirit of 'maximum entropy' Bayesian analysis (Jaynes 1988) may be formed by:

$$c_{\text{true}}^{i,j<M,1} \sim \text{Uniform}(Z_{\geq 0}) \text{ with}$$

$$c_{\text{true}}^{i,j<M,2} = \sum_{w:MB_w \in SA1_i} c_{\text{true}}^{w,j<M,1} \text{ and } c_{\text{true}}^{i,j<M,3} = \sum_{w:MB_w \in SA2_i} c_{\text{true}}^{w,j<M,1}$$

$$\text{and } c_{\text{true}}^{i,M,k} = \sum_{w<M} c_{\text{true}}^{i,w,k}.$$

That is, we suppose the Mesh Block count in any category and any cell has a (improper) uniform probability of being any non-negative integer, with the aggregate and row counts defined from summation over these Mesh Block values. Once conditioned on the available data through the likelihood function, this model admits a proper Bayesian posterior, which we will refer to here as the maximum entropy solution.

In geospatial statistical analyses it is common to introduce hierarchical priors that leverage existing information or theoretical expectations concerning the correlation of outcomes with spatially varying covariates and the auto-correlation of outcomes across nearby spatial units (Bhatt, Weiss et al. 2015, Held, Hens et al. 2019). We propose that priors of this form can be readily and effectively used to tune TableBuilder count reconstructions to the purpose of a given scientific analysis. For ease of explanation we will again consider a concrete example.

Suppose that one is interested in reconstructing SA1 counts of one of the $M - 1$ classes under consideration (here: "Multiple Family Household") and that the Index of Relative Socio-Economic Advantage and Disadvantage (IRSAD) decile is proposed as a covariate for explaining the proportion of members of that class in each area. To this we add a spatial random effect with exponential covariance function and nugget (Diggle, Tawn et al. 1998) defined using the centroids of the SA1 units. Combining this classic geospatial model with a maximum entropy style prior over the row totals at SA1 level gives a partial prior:

$$c_{\text{true}}^{i,M,2} \sim \text{Uniform}(Z_{\geq 0}) \text{ and } c_{\text{true}}^{i,1,2} \sim \text{Binomial}(\phi_i, c_{\text{true}}^{i,M,2})$$

$$\text{with logit}(\phi_i) \sim \beta[\text{IRSAD}_i] + g_i + \epsilon_i$$

$\beta_{1:10} \sim \text{Normal}(0,1)$, $g_{(\cdot)} \sim \text{ExpGaussianProcess}(\Phi)$, $\epsilon_i \sim \text{Normal}(0, \sigma_{\text{iid}}^2)$.

This prior may be completed by adding suitable hyper-priors on the hyper-parameters of the random effects distribution (Diggle, Tawn et al. 1998), and then supposing non-informative, maximum entropy style priors on the allocations of the remaining SA1 level class counts and then of all class counts to the lower Mesh Block units. The SA2 level counts are completed by summation as for the original maximum entropy model.

*Markov Chain Monte Carlo Algorithm*

The posterior of the above Bayesian model is intractable to closed form analytic evaluation, meaning that it is necessary to seek a computational posterior approximation. While many posterior approximation problems may be readily solved through an 'off-the-shelf' sampling codes (such as Stan or JAGS), the above models feature several challenges for efficient representation in standard probabilistic programming languages. First, the handling of structural zeros (i.e., true zero cell counts) and the production of random zeros through noise addition and suppression creates a 'mixture model' term in the likelihood function that risks numerical overflows. Second, the nested hierarchy of row total and spatial aggregation constraints places strict 'geometric' bounds on the support of the posterior parameter space. And third, the parameter space is discrete (i.e., non-negative integer valued counts), which rules out gradient based sampling methods (such as hybrid Monte Carlo) designed for continuous random variables. Given these challenges we instead develop our own Markov Chain Monte Carlo procedure.

For both the maximum entropy model and the geostatistical model we propose a new set of parameters at the Mesh Block level, $\{c_{\text{true}}^{i,j<M,1}\}^{\text{proposed}}$, by adding or subtracting a single count from a randomly chosen category in a randomly chosen location. Each corresponding $\{c_{\text{true}}^{i,M,1}\}^{\text{proposed}}$, $\{c_{\text{true}}^{i,j<M,2}\}^{\text{proposed}}$, $\{c_{\text{true}}^{i,M,2}\}^{\text{proposed}}$, $\{c_{\text{true}}^{i,j<M,3}\}^{\text{proposed}}$ and $\{c_{\text{true}}^{i,j,3}\}^{\text{proposed}}$ element for that proposal (i.e., the implied proposals of row sums and new counts by datum over the hierarchy of areal units) are then computed. The proposal is accepted or rejected according to a Metropolis-Hasting step considering the ratio of likelihoods for the proposed and current count tables.

In the geostatistical model we must augment the Metropolis-Hasting allocation step at SA1 level with the geostatistical prior. This is achieved by splitting the algorithm into a blocked Gibbs sampler (Roberts and Sahu 1997), whereby first the geostatistical regression model is fitted to the current $\{c_{\text{true}}^{i,1,2}\}^{\text{current}}$ and $\{c_{\text{true}}^{i,M,2}\}^{\text{current}}$ counts. A draw of the regression coefficients, hyper-parameters and the spatial random field are made to provide a conditional update against this prior, with which the model is advanced using the Metropolis-Hastings steps as above augmented with prior weights. For computational efficiency the empirical logit approximation of the binomial likelihood may be used, allowing a Multivariate Normal approximation of the geostatistical model. The experienced computational statistician may find that the full implementation scheme for this posterior sampler is most readily understood through inspection of the standard R code included here in the Supplementary Information.

*Posterior Predictive Checks*

Posterior predictive checks have been advocated in the Bayesian literature as a useful tool for diagnosing issues of model misspecification (Gelman, Meng et al. 1996). Posterior predictive statistics may include distributions of the characteristics of mock datasets generated under the fitted model. By comparing these against the same statistic computed on the real data it is possible to identify discrepancies that may indicate an inflexibility or other undesirable feature of the proposed model. Two posterior predictive statistics useful for model checking here are as follows:

1. the median and other quantiles of the difference between the aggregated TableBuilder (or model-perturbed) output Mesh Block counts and the enclosing TableBuilder (or model-perturbed) SA1 counts; and

2. the median and other quantiles of the difference between the row sums of TableBuilder (or model-perturbed) outputs over Mesh Block classes and the TableBuilder (or model-perturbed) row totals.

Both statistics are proposed to shed light on the suitability of the proposed perturbation model to represent the true ABS perturbation procedure. As with any posterior predictive checks, a failure to identify mismatches between the model and real data statistics does not confirm the validity of the model, which in any case is assumed (philosophically) to be inherently misspecified. Nevertheless, it is hoped that these statistics will identify any gross mismatches between the model and the true data generating process, feeding information forwards for future cycles of model refinement as required.

*Empirical Coverage*

Another tool to investigate the behaviour of Bayesian models is the comparison of posterior distributions against the ground truth when fitting with mock datasets for which the latter is known exactly (unlike in a real-world analysis). Here we focus on the metric of Bayesian 'coverage', examining what proportion of true latent count values are contained within our posterior 95% credible intervals for the cell counts at each areal level. While Bayesian credible intervals are not guaranteed to deliver Frequentist style performance, meaning that they would enclose at least the nominal fraction of true values, a large divergence in behaviour is generally a sign of an unusual (perhaps even pathological) model that should be treated with caution (or an incorrectly implemented posterior simulation code).

*Mock Datasets*

To generate mock datasets we begin by requesting total ABS TableBuilder dwelling counts (perturbed) for the Perth Greater Capital City region at the scale of Mesh Blocks,

SA1 and SA2 areas. We reconcile the SA1 and Mesh Block totals with the enclosing SA2 totals by adding (or, less commonly, subtracting) counts as required, choosing randomly from amongst the finer enclosed units until equality is reached. A random value of the IRSAD covariate effect and a spatial random field are then drawn according to the geostatistical model given above, and a fiducial set of true latent SA1 class types for the (generally) rarest member of an imaginary four class category are generated by binomial sampling. The remaining fiducial counts for the other two classes at SA1 level are generated by drawing from a beta-binomial distribution. Corresponding Mesh Block fiducial values are then created in proportion to the Mesh Block population totals, and the SA2 fiducial values follow again by summation.

Mock outputs from our minimal perturbation model are then easily created by sampling from a truncated discrete Normal distribution for all non-zero cells and applying a subsequent suppression of all noise-added counts of 1 or 2 to zero. Analysis of posteriors constructed from mock datasets generated with this procedure constitutes our well-specified model scenario. Mock outputs from our individual record perturbation model are generated by first creating a mock record set equal to the total dwelling count in Perth GCC and then assigning these records at random to fill up the fiducial cell counts at Mesh Block level. Analysis of posteriors constructed from mock datasets generated with this procedure constitute our misspecified scenario.

**Results**

*Multi-scale Bayesian Reconstructions*

In Figure 1 we present a series of maps for a section of central Perth showing the posterior mean estimates of the number of "Multiple Family Household" dwellings per Mesh Block, SA1 and SA2 unit, as reconstructed with our geostatistical model. The

TableBuilder outputs that served as inputs to this model are shown below the reconstruction at each areal scale for reference. The impact of low count suppression on the TableBuilder outputs at fine areal scales is readily observed in the sparsity of these maps at Mesh Block and SA1 levels, as is the impact of the model in attempting to probabilistically redistribute the counts known to be missing from consideration of the SA2 totals. Another effect of the model that is also readily apparent is the smoothing of high values from the TableBuilder outputs: under the assumed perturbation model, a cell count that survives the thresholding step is more likely to have benefitted from a positive random perturbation than a negative one.

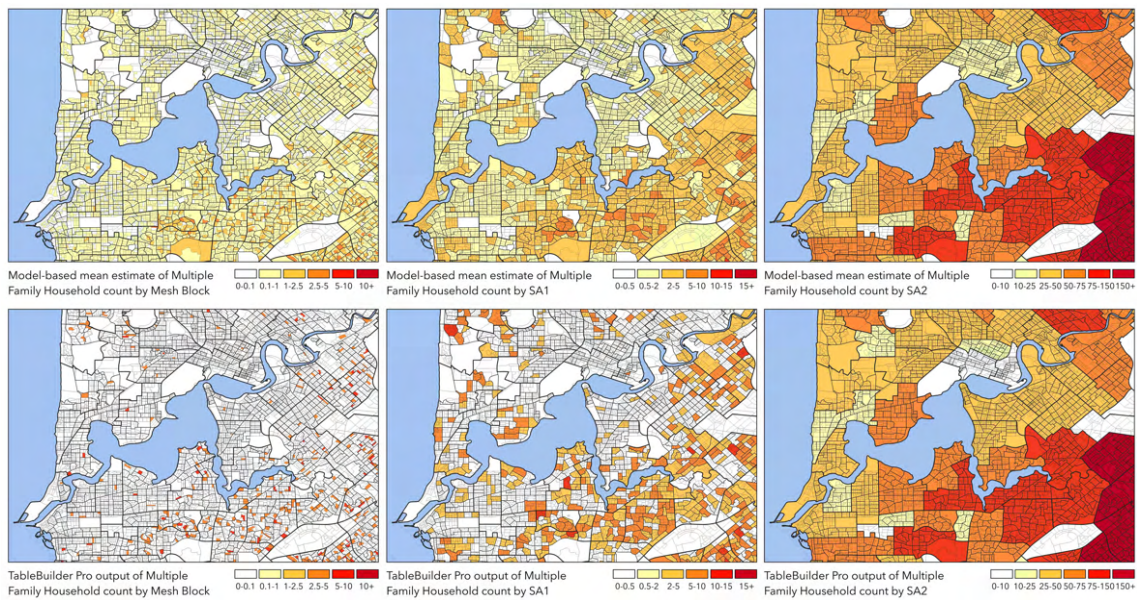

Figure 1: Model-based reconstructions of the spatial pattern across central Perth of dwellings classified as containing "Multiple Family Households" (*top row*) compared against the perturbed input data from TableBuilder (*bottom row*). The color scales for each areal unit (from Mesh Block on the *left* to SA1 in the *middle* and SA2 on the *right*) have each been adjusted to highlight the underlying patterns, but have been kept consist on each row for direct comparison.

Importantly, since this is a fully statistical reconstruction approach, samples from the posterior and posterior summaries are also readily produced. In Figure 2 we present two posterior samples as maps of the "Multiple Family Household" dwelling count at Mesh Block level, each representing one of many possible worlds consistent with the available data. In many areas the presence of at least one "Multiple Family Household" in the enclosing SA1 unit is indicated by the data, but the model remains uncertain as to which Mesh Block it belongs; while in other areas the model is confident from the available data that a specific mesh block is host to "Multiple Family Household" dwellings or alternatively that there are none in that area. Importantly, in each of these possible worlds the sums of dwelling type counts across areal units will be both internally consistent and faithful to the input data. We also show for reference in Figure 2 the width of the 'pointwise' posterior 95% credible interval at Mesh Block level. Here we see that the typical 95% credible interval on any Mesh Block is in the 2-5 counts range, which reflects the scale of the random noise in the error model assumed here. To achieve stronger posterior bounds one would need to introduce a substantial amount of additional information beyond the IRSAD deciles used in our geostatistical model. Examples for this application might include maps of development zone categories, building heights or footprint areas, or other such measures likely relevant to the probability of a dwelling being classified as a "Multiple Family Household".

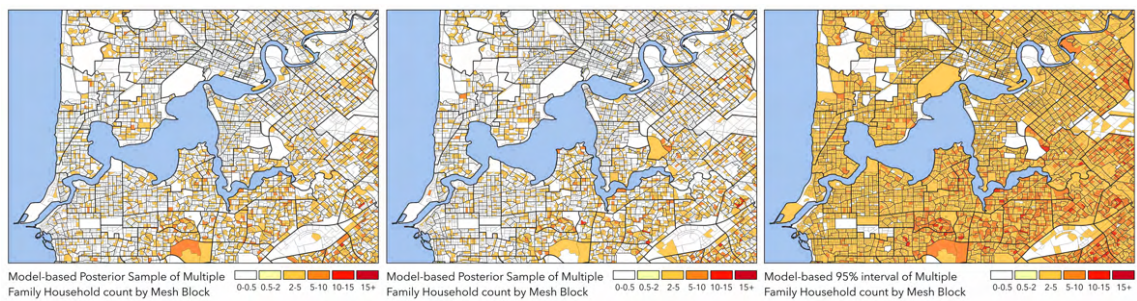

Figure 2: Two posterior samples of the "Multiple Family Household" count by Mesh

Block for central Perth (*left & middle*), and the width of the 95% posterior credible interval over all posterior samples (*right*).

In Table 1 we present posterior summaries of the fitted parameters in the geostatistical model concerning the relationship between IRSAD and the proportion of "Multiple Family Household" dwellings. Of course, for these uncertainty intervals to have value we must establish evidence that the model is likely to be well-performing despite the known misspecification in our chosen likelihood function (i.e., our proposed minimal perturbation model). Addressing this challenge will be the focus of the subsequent two sub-sections of the Results present here; first though, we will examine the difference in reconstructions between our geostatistical model and our maximum entropy model.

| IRSAD Decile: | 1 | 2 | 3 | 4 | 5 | 6 | 7 | 8 | 9 | 10 |
|---|---|---|---|---|---|---|---|---|---|---|
| *Lower (2.5%)* | -0.76 | -0.88 | -0.99 | -1.29 | -0.96 | -0.88 | -0.85 | -0.82 | -0.30 | (ref) |
| *Median (50%)* | -0.67 | -0.78 | -0.86 | -1.12 | -0.85 | -0.78 | -0.72 | -0.70 | -0.13 | (ref) |
| *Upper (97.5%)* | -0.60 | -0.66 | -0.77 | -0.97 | -0.74 | -0.68 | -0.61 | -0.59 | 0.04 | (ref) |

Table 1: Posterior estimates (median and 95% credible interval) for the association between IRSAD and the proportion of dwellings per SA1 area identified as "Multiple Family Households". The effects here are additive to the logit transformed proportion with the least disadvantaged and most advantaged decile (10) serving as the reference (i.e., zero effect).

In Figure 3 we present a selection of maps for comparing the reconstruction of the "Multiple Family Household" dwelling count at SA1 level under our maximum entropy model with that from our geostatistical model. First, we show the posterior mean count from the former for direct comparison against that shown in Figure 1. Broadly speaking, there is little noticeable difference immediately apparent in the posterior mean count totals at this scale. However, we also include here the posterior mean estimates of the proportion of "Multiple Family Household" dwellings at SA1 level under both the maximum entropy and geostatistical models. Since the latter specifically includes information on the IRSAD decile and a spatial smoothing term acting on this proportion in its prior structure, we anticipated this to be a key area of difference in outputs from the two reconstruction models.

Indeed it is immediately clear from inspection of Figure 3 that the geostatistical model has massively reduced the proportion of SA1 cells in central Perth assigned high fractions of "Multiple Family Household" dwellings in comparison with the maximum entropy version. Many of these areas with very high fractions in the latter case will be recognised by those familiar with Perth city as areas with low total dwelling counts, such as the Wembley and Royal Perth Golf Clubs. However, non-trivial examples of more gentle shrinkage against high fractions in the geostatistical map are also seen on both sides of the river.

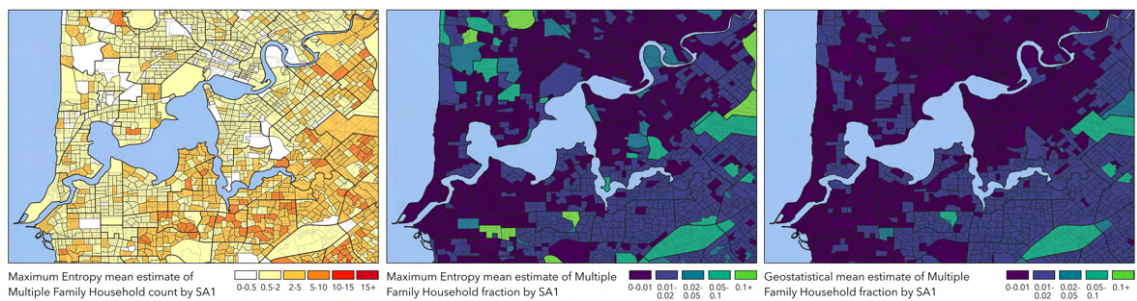

Figure 3: Illustration of the difference between the reconstructions recovered under our maximum entropy and geostatistical models for the spatial pattern at SA1 level across central Perth of dwellings classified as containing "Multiple Family Households". The posterior mean counts under the maximum entropy model are shown in the *left* column, for direct comparison against the geostatistical equivalent shown in Figure 1. The posterior mean fraction of "Multiple Family Household" dwellings per SA1 unit are also shown for the maximum entropy model (*middle*) and geostatistical model (*right*).

*Performance of Posterior Predictive Checks*

In Tables 2 and 3 we present the results of our posterior predictive checking analyses derived from our Bayesian posteriors recovered against the real "Multiple Family Household" outputs from TableBuilder, as well as those from mock datasets created under the well-specified and misspecified scenarios. The first posterior predictive check concerns the difference between summed Mesh Block cell counts in their enclosing SA1 units and those queried directly at the SA1 level. This probes our representation of the under-counting problem noted in earlier analyses. The second posterior predictive check concerns the difference between the row-summed Mesh Block cell counts and their directly queried row totals. This probes our representation of the perturbation process at the smallest cell counts.

| *Quantiles:* | 2.5% | 25% | 50% | 75% | 97.5% |
|---|---|---|---|---|---|
| *Real data* | -6 [-5,-5] | -3 [-3,-1] | 0 [0,0] | 0 [0,0] | 3 [4,4] |
| *Mock data: well-specified* | -9 [-8,-7] | -3 [-4,-3] | -1 [-1,1] | 1 [1,1] | 6 [4,5] |
| *Mock data: misspecified* | -10 [-7,-7] | -4 [-3,-3] | -2 [-1,0] | 1 [1,2] | 6 [5,5] |

Table 2: Results for our first posterior predictive check examining the posterior predictive distribution of the difference between perturbed "Multiple Family Household" counts

aggregated from Mesh Blocks to SA1 units against those queried directly at the SA1 level. The value at each of 5 quantiles seen in the dataset considered is compared with the 95% posterior predictive interval (in square brackets) in each case.

| Quantiles: | 2.5% | 25% | 50% | 75% | 97.5% |
|---|---|---|---|---|---|
| *Real data* | -7 [-5,-4] | -1 [-3,-1] | 0 [0,0] | 2 [2,2] | 8 [4,4] |
| *Mock data: well-specified* | -5 [-4,-4] | -1 [-1,-1] | 0 [0,0] | 2 [2,3] | 6 [5,5] |
| *Mock data: misspecified* | -6 [-4,-4] | 0 [-1,-1] | 0 [0,0] | 2 [2,3] | 8 [5,5] |

Table 3: Results for our second posterior predictive check examining the posterior predictive distribution of the difference between perturbed "Multiple Family Household" counts aggregated row-wise in Mesh Blocks against those queried directly for the row sum. The value at each of 5 quantiles seen in the dataset considered is compared with the 95% posterior predictive interval (in square brackets) in each case.

For readers unfamiliar with posterior predictive checks, it may be surprising to see that even in the well-specified scenario (where our mock data are drawn from exactly the same generating process as assumed by the Bayesian model) there are a number of places where the data statistic falls outside of the posterior predictive 95% interval. Although this could occur by chance under a well-calibrated test, there is in fact no expectation that posterior predictive checks will be well-calibrated (in terms of having a uniform distribution under the null; Gelman (2013)). Nevertheless, if the statistics chosen for the test are meaningful they can still provide useful insights into model suitability (Gelman, Meng et al. 1996).

Here we see that in the first posterior predictive check the real data and well specified models show similar levels of discrepancy between their data realisations and the posterior predictive intervals, while data in the misspecified case is far from the interval at the lower extreme of the posterior predictive distribution. That is, under our individual record perturbation model we find a larger range of differences upon aggregation of the mock data at Mesh Block level to SA1 level than we do under our Bayesian analysis with the (more tractable) minimal perturbation model. We suggest that this extra variation in the individual version is due to the auto-correlation between output counts in cells containing common individual members inherent in its construction. If so, it may be the case that auto-correlation induced by the real TableBuilder perturbation is not as impactful as in the individual records model we have proposed here.

In the second posterior predictive check both the real data and the misspecified data show extreme discrepancies against the tails of their posterior predictive distributions. That is, the true perturbation model being applied within TableBuilder seems to be producing a wider spread of differences between the row-sums of perturbed counts within Mesh Blocks against direct query of the row totals than results from our minimal perturbation model. Our individual record perturbation model produces a similar level of discrepancy which suggests that in this regard the nature of the misspecification of the former against our minimal perturbation model is similar to that of the true model. This gives confidence that the individual record perturbation model is a suitable tool for exploring the impacts of misspecification in our experiments.

*Empirical Coverage of Credible Intervals*

Examining the Frequentist style behaviour of Bayesian models when fitting against simulated (mock) datasets (with known fiducial values for the latent variables) is another useful diagnostic of model performance. When fitting mock datasets simulated under the same generating process as assumed in the Bayesian model we anticipate that the credible intervals will provide coverage close to their nominal values. That is, the 95% credible interval will contain the fiducial value in at least 95% of independent data realisations, regardless of the true parameter values from which data are simulated. Here the Frequentist coverage statement would be made for a single estimated variable, such as the dwelling count in a single specific Mesh Block.

For geospatial analyses of the nature considered here we propose that a more interesting characteristic is the empirical coverage fraction calculated over the set of all spatial units of a given size in the map under investigation. This statistic speaks directly to the average performance of the model over the areal units in a single fitting experiment against a specific realisation of the underlying spatial random field. As a result, useful insights can often be gained by visual comparisons of areas having adequate and inadequate empirical coverage against the underlying field values and the mock data values for that realisation. Multiple realisations of this global coverage analysis are required to then probe the insights further, but rarely will adequate computational time and power be available from which to build accurate long-run Frequentist coverage estimates of the traditional definition.

Over five realisations of the spatial field and mock dataset construction we find mean global coverage fractions of 0.95 at Mesh Block level, 0.94 at SA1 level and 0.90 at SA2 level for our 95% credible intervals in the well specified scenario under our geostatistical model. These are all sufficiently close to the nominal value of 0.95 that we are reassured both that our Markov Chain Monte Carlo sampler is likely performing correctly and that the model does not suffer any unusual pathologies. More interesting is the misspecified case, for which we find equivalent mean global coverage fractions of 0.94 at Mesh Block level, 0.92 at SA1 level and 0.90 at SA2 level. A decline in coverage is an often-encountered problem for Bayesian inference under misspecification, yet the reductions here are very small. As such, if our individual record perturbation model was a faithful representation of the true perturbation model applied in TableBuilder we would expect negligible consequences from its approximation by our minimal perturbation model in the likelihood function.

Examples of the spatial pattern of coverage of the fiducial counts at SA1 level for both the well-specified and misspecified scenarios are shown in Figure 4. These maps cover a region of north-east Perth with a wide dynamic range of fiducial "Multiple Family Household" dwelling counts in the simulated dataset. Areas of coverage failure in both scenarios are generally spatially clustered. This is likely an effect of the nested data and model structure for the spatial units concerned; areas where the mock TableBuilder SA2 count is significantly higher or lower than the fiducial value would naturally be harder to learn well under any model.

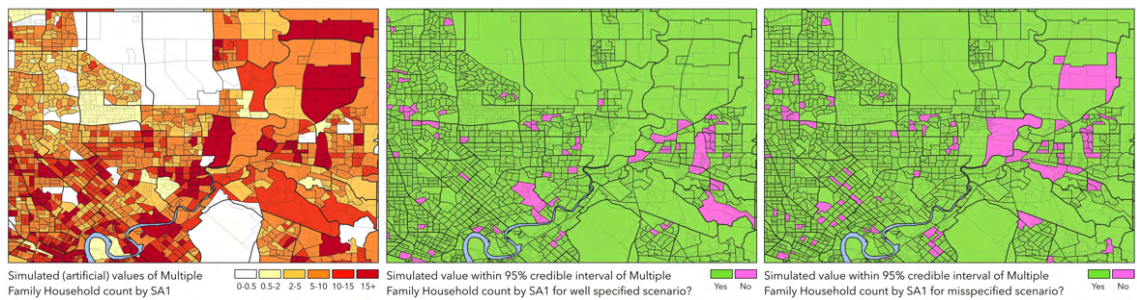

Figure 4: Investigation of empirical coverage for our geostatistical model of count reconstruction using mock datasets with the true latent values shown in the *left* column; results are shown both for mock TableBuilder outputs produced in the well-specified scenario (i.e., with our minimal perturbation model; *middle*) and in the misspecified scenario (i.e., with our individual record perturbation model; *right*).

**Discussion**

In their *Australian Geographer* paper identifying the challenges of working with ABS TableBuilder outputs, Kok, Tuson et al. (2022) highlight the importance of disease maps for the estimation of disease burden and resources allocation by reference to the work of the Malaria Atlas Project (Hay, Noor et al. 2005, Hay, Battle et al. 2013). Interestingly, in 2020 MAP relocated from the University of Oxford to the Telethon Kids Institute in Perth, Western Australia, adding to the strengths of the Australian infectious disease mapping community. A guiding principle of MAP's work has always been that when working with disease data a statistical approach is essential to ensure that the uncertainties attached to disease maps can be communicated clearly to the intended audience of policy makers, researchers and the general public. Importantly, Hay and Snow (2006) note that:

*"A very considerable research effort is also required to evaluate those statistical techniques needed to relate the PR [prevalence] and environmental data for extensive map predictions with confidence intervals."*

That is, the challenge for researchers working in this field is not only to develop novel statistical methods for disease mapping, but also to subject those methods to scrutiny to ensure that their performance can be monitored and improved over time.

In this study we have demonstrated that a statistical treatment of the ABS TableBuilder outputs is both feasible and effective using a likelihood-based, Bayesian methodology. Specifically, we have demonstrated that under a simple approximation for the incompletely-known perturbation procedure that is applied to the true cells counts within TableBuilder, both maximum entropy style and geostatistical Bayesian models may be fitted with posterior sampling achieved through Markov Chain Monte Carlo. Moreover, we have interrogated the performance of these models in both the well-specified regime (i.e., with mock data simulated under the assumed perturbation model) and in the misspecified regime (i.e., with mock data simulated under a more complicated alternative perturbation model). The performance of our model was satisfactory in both instances and we further demonstrated through posterior predictive checks that our individual records perturbation model is produced similar impacts to the true process.

Further work is warranted in a number of possible research directions to build on this foundation. A range of alternative error models might be explored to determine whether they can achieve new insights into the possible impact of the true ABS TableBulder perturbation algorithm on the output cell counts and the accuracy of the posterior reconstructions. If the models so identified are intractable to likelihood-based analysis, as is the individual record perturbation model proposed for the misspecified scenario in this study, then attention might be given to how best to then proceed with Bayesian analyses. 'Likelihood-free' to inference, including Approximate Bayesian Computation (Lintusaari, Gutmann et al. 2017) and Bayesian Synthetic Likelihood (Price, Drovandi et al. 2018), propose simulation-based approach to inference with intractable likelihoods, although these methods are not typically feasible in very large parameter spaces. Methods for improving uncertainty estimates for misspecified Bayesian models are also being actively investigated (Miller and Dunson 2018, Lyddon, Holmes et al. 2019, Huggins and Miller 2023), although the auto-correlated nature of spatial models is beyond the scope of these current advances.

The ability of our proposed method to produce realisations of fine scale cell count maps for arbitrary TableBuilder classes should also motivate further work of more applied nature. Bayesian posterior sampling provides many realisations of the possible fine scale cell count map that constitute a collection of 'possible worlds' representing our beliefs about the range of likely maps given the available data. From each realisation one may construct new products, such as maps of "hot spot" locations (as relevant, for example, to the 'gayborhood' analysis of Callander, Mooney-Somers et al. (2020)); the collection of which then represents the Bayesian posterior of that new product.  A researcher interested in the comparing a measure like 'mean household distance to a major road' for a particular cohort (such as children with long-term health conditions) between SA1 units could compute this measure on each Mesh Block realisation to create the range of possible values at each SA1.  Downstream analyses with these distributions will then require an 'errors-in-variables' style analysis, which is more complicated than a simple regression but well within the capability of standard statistical modelling packages (such as R and SPSS).

Datasets subject to perturbations introduced (either in the spirit of, or directly informed) against differential privacy policies are becoming increasingly available and increasingly important for geographical analyses.  The Demographic and Health Information survey program, which delivers national datasets crucial to the understanding of health and development in low and middle income countries, introduces a randomisation to the reported cluster coordinates to preserve the privacy of respondent communities (Allshouse, Fitch et al. 2010).  Both Facebook and Google have released human movement datasets constructed from aggregate user movement counts adjusted with noise-injection and small count suppression. Analyses with these datasets, including an investigation of the hierarchy of intra-urban mobility (Bassolas, Barbosa-Filho et al. 2019) and the travel of potential disease carriers between countries (Shepherd, Atherden et al. 2021), have worked around the suppression problem by considering only origin-destination pairs with non-zero counts.

In these differential privacy examples the perturbation algorithms are more clearly described in documentation (owing to their greater simplicity) than that employed in

TableBuilder. Nevertheless, modellers are still not allowed the totality of information they might like to reconstruct the process in each case. For instance, the population sampling frames from which the DHS clusters are selected are not typically made public, and the scaling of Google movement values is also opaque to users. Kok, Tuson et al. (2022) conclude with a strong call for the ABS to return the additivity step from pre-2016 TableBuilder to their perturbation algorithm. Instead, we would conclude with an appeal to the ABS for more information on the perturbation process, or for the provision of mock datasets created under the same perturbation process alongside known fiducial cell counts that researchers could use to test the fidelity of their reconstruction procedures. As noted in the Introduction by reference to Rinott, O'Keefe et al. (2018), differential privacy procedures should be robust to outside knowledge of the perturbation process, and the probabilistic reconstruction of the original aggregate data for research purposes is a fair ambition of the end-user.

**Conclusions**

The perturbation of true cell counts from the Australian Census before distribution to researchers as ABS Tablebuilder outputs is an important step to preserving the privacy of individual records. We have demonstrated that for researchers working with this perturbed data it is feasible and effective to create statistical reconstructions that achieve consistency in aggregation across multiple spatial scales. Misspecification between the true perturbation algorithm and the minimal model for tractable inference proposed here is identified via posterior predictive checks. Adequacy of our Bayesian model is demonstrated under some level of misspecification, although the impact of the true perturbation process on the posterior remains uncertain.